\documentclass[12pt]{article}

\usepackage{amsfonts}
\usepackage[mathscr]{eucal}
\usepackage{bm}
\usepackage{cite}

\newcommand{\be}{\begin{equation}}
\newcommand{\ee}{\end{equation}}
\newcommand{\bea}{\begin{eqnarray}}
\newcommand{\eea}{\end{eqnarray}}

\newcommand{\p}[1]{(\ref{#1})}

\newcommand{\+}{\dagger}

\begin{document}

\begin{titlepage}

\vspace*{1.5cm}

\renewcommand{\thefootnote}{\dag}
\begin{center}

{\LARGE\bf $\mathcal{N}{=}\,2$ supersymmetric hyperbolic }

\vspace{0.45cm}

{\LARGE\bf Calogero-Sutherland model }

\vspace{1.5cm}

{\large\bf Sergey Fedoruk}
 \vspace{0.5cm}

{\it Bogoliubov Laboratory of Theoretical Physics, }\\
{\it Joint Institute for Nuclear Research,}\\
{\it 141980 Dubna, Moscow region, Russia} \\
\vspace{0.1cm}

{\tt fedoruk@theor.jinr.ru}

\vspace{1.5cm}

\end{center}

\vspace{1.2cm} \vskip 0.6truecm \nopagebreak

\begin{abstract}
\noindent
\qquad The $\mathcal{N}{=}\,2$ supersymmetric hyperbolic Calogero-Sutherland model obtained in arXiv:1902.08023 by gauging the $\mathcal{N}{=}\,2$ superfield matrix system is studied. Classical and quantum $\mathcal{N}{=}\,2$ supersymmetry generators are found. The difference in the structure of classical and quantum supercharges is established. It is shown that, unlike classical supercharges, quantum supersymmetry generators can be limited to an invariant sub-sector that does not contain off-diagonal fermion operators. The Lax pair for supersymmetric generalization of the hyperbolic Calogero-Sutherland system is constructed.
\end{abstract}

\vspace{3cm}


\newpage

\end{titlepage}

\setcounter{footnote}{0}
\setcounter{equation}0
\section{Introduction}

Since the multi-particle integrable Calogero-Sutherland systems \cite{C,Su,CalRM}   (see \cite{OP,Per,Poly-rev} for reviews)
hold a special place in modern theoretical physics, numerous attempts are made to obtain various generalizations of these models.
Supersymmetric generalizations of the Calogero-Sutherland models are of particular interest among possible developments.
Unlike fairly well-developed ${\mathcal N}$-extended supersymmetric versions of the rational Calogero model
\cite{FrM,BHKVas,BGK,BGL,Wyl,GLP-2007,FIL08,Fed-2010,KL-10,FI,FILS,KLS-18,KLS-18b,Feig,KLS-19} (see, for example, the review \cite{superc}),
supersymmetric generalizations of the Calogero-Sutherland systems \cite{Su,CalRM} are understood to date rather badly
(see for example
\cite{SSuth,BrinkTurW,BorManSas,IoffeNee,DeLaMa,Serg,SergVes,Feig,Pr-2019,Kr-19,KLS-19,KL-20} and references therein).

In a recent paper \cite{FIL19}, $\mathcal{N}{=}\, 2$ and $\mathcal{N}{=}\, 4$ supersymmetric generalizations
of the multi-particle hyperbolic Calogero-Sutherland system \cite{C,Su}  (see \cite{OP,Per,Poly-rev} for reviews)  were constructed.
These systems were derived from the matrix one-dimensional superfield systems by the gauging procedure \cite{DI-06-1}.
This method is a direct generalization of the gauging procedure used in \cite{FIL08} (see also \cite{FIL09,Fed-2010,superc})
for obtaining ${\mathcal N}{=}\,1,2,4$ supersymmetric extensions of the rational Calogero models.\footnote{
Rational Calogero-Moser systems with deformed supersymmetry were derived by the gauging procedure in \cite{FI,FILS}.
Similar nongauging matrix systems with an extended set of fermionic fields were considered in \cite{KLS-18,KLS-18b} and \cite{FI-18}.
The matrix description of the Calogero models was also considered in \cite{Poly-gauge,Gorsky,FeKl,Poly-rev}.}

In the $\mathcal{N}{=}\, 2$ matrix model considered in \cite{FIL19}, after putting the Wess-Zumino gauge and elimination of auxiliary fields,
the $n$-particle system is described by the on-shell component action
$
{\displaystyle S_{\rm matrix} = \int \mathrm{d}t \, L_{\rm matrix} }
$
with the Lagrangian
\begin{eqnarray}
L_{\rm matrix}  &  = & \frac12\,{\rm Tr}\Big( \,
X^{-1}\nabla\! X \,X^{-1}\nabla\! X\Big)  +  \frac{i}{2}\,{\rm Tr} \Big( X^{-1}\bar\Psi X^{-1}\nabla \Psi - X^{-1}\nabla \bar\Psi X^{-1}\Psi \Big)
\nonumber\\ [5pt]
&&
+ \frac{i}{2}\, \Big(\bar Z \nabla\! Z - \nabla\! \bar Z Z\Big)
- \, \frac{1}{4}\,{\rm Tr} \Big( X^{-1}\bar\Psi X^{-1}\bar\Psi X^{-1}\Psi X^{-1}\Psi \Big)
+ c\,{\rm Tr} A\,,
\label{N2Cal-com}
\end{eqnarray}
which involves the following matrix fields ($a,b=1,\ldots ,n$):
\begin{itemize}
\item
the positive definite Hermitian $c$-number $(n{\times}n)$--matrix field
$$
X(t):=\|X_a{}^b(t)\|\,, \qquad ({X_a{}^b})^* =X_b{}^a \quad (X^\+=X)\,,\qquad \det X \neq 0\,,
$$
\item
the complex $n{\times}n$--matrix fields with Grassmannian elements
$$
\Psi(t):=\|\Psi_a{}^b(t)\|\,, \qquad \bar\Psi(t):=\|\bar\Psi_a{}^b(t)\|\,, \qquad ({\Psi_a{}^b})^* =\bar\Psi_b{}^a \quad (\Psi^\+=\bar\Psi)\,,
$$
\item
the complex $c$-number $\mathrm{U}(n)$-spinor fields ($(1{\times}n)$-- and $(n{\times}1)$--matrices)
$$
Z(t):=\|Z_a(t)\|\,, \qquad \bar Z(t):=\|\bar Z^a(t)\|\,, \qquad \bar Z^a = ({Z_a})^*\,.
$$
\item
$n^2$ gauge fields that form the Hermitian $c$-number $(n{\times}n)$--matrix field
$$
A(t):=\|A_a{}^b(t)\| \,, \qquad ({A_a{}^b})^* =A_b{}^a \quad (A^\+=A)\,,
$$
and are present in the last term (the Fayet-Iliopoulos term) and in the definition of the
covariant derivatives
\begin{equation}\label{cov-der-xp}
\nabla\! X = \dot X +i\, [A, X]\,, \qquad \nabla \Psi = \dot \Psi +i\, [A,\Psi]\,, \qquad \nabla \bar\Psi = \dot {\bar\Psi} +i\, [A,\bar\Psi]\,,
\end{equation}
\begin{equation}\label{cov-der-z}
\nabla\! Z = \dot Z + iAZ\,, \qquad \nabla\! \bar Z = \dot{\bar Z} -i\bar Z A\,.
\end{equation}
\end{itemize}
In the Fayet-Iliopoulos term the quantity $c$ is a real constant.

In this  paper, a detailed study of the many-particle system \p{N2Cal-com} is carried out.
In particular, classical and quantum supercharges are found and the Lax pair of the considered system is constructed.

The plan of the paper is as follows.
In Section~2, it is shown, the system \p{N2Cal-com}  with a completely fixed gauge is exactly
the multi-particle hyperbolic Calogero-Sutherland system \cite{C,Su} in the boson limit.
In Section~3, the Hamiltonian formulations of the matrix system and the reduced system,
obtained by eliminating purely gauge bosonic off-diagonal matrix fields, are constructed.
This allowed in Section~4 to find, using the Noether procedure, a full set of $\mathcal{N}{=}\, 2$ supersymmetry generators.
In Section~5, the Lax pair for the supersymmetric generalization of the hyperbolic Calogero-Sutherland system under consideration is constructed.
Section~6 is devoted to the construction of quantum $\mathcal{N}{=}\, 2$ supersymmetry generators for the hyperbolic Calogero-Sutherland system.
It is shown here that, unlike classical supercharges, quantum supersymmetry generators can be limited to an invariant sub-sector that does not contain off-diagonal fermion operators.
The last Section~7 contains a summary and outlook.

\setcounter{equation}0
\section{Lagrangian consideration}

The Lagrangian (\ref{N2Cal-com}) is invariant, up to the total derivative, with respect to the local $\mathrm{U}(n)$
transformations, $g(\tau )\in \mathrm{U}(n)$,
\begin{equation}\label{Un-tran}
X \rightarrow \, g X g^\+ \,, \qquad  Z \rightarrow \, g Z \,, \quad \bar Z
\rightarrow \,
\bar Z g^\+\,,
\qquad
A \rightarrow \, g A g^\+ +i \dot g g^\+\,.
\end{equation}
\begin{equation}\label{Un-tran-Psi}
\Psi \rightarrow \, g \Psi g^\+ \,, \qquad  \bar\Psi \rightarrow \, g \bar\Psi g^\+\,.
\end{equation}
The allowed gauge for the local transformations (\ref{Un-tran}) is the following:
\begin{equation}\label{X-fix}
X_a{}^b =0\,,\qquad a\neq b\,,
\end{equation}
i. e.
\begin{equation}\label{X-fix-com}
X_a{}^b = x_a \delta_a{}^b \,.
\end{equation}
As a result, the Lagrangian (\ref{N2Cal-com}) becomes
\begin{eqnarray}\nonumber
L \!\!&\!=\!&\!\!  \frac12\, \sum_{a,b} \left[ \frac{\dot x_a \dot x_a}{(x_a)^2}
+ i \left(\bar Z^a \dot Z_a - \dot{\bar Z}{}^a Z_a\right) +
\frac{i}{x_a x_b}\left(\bar\Psi_a{}^b \dot\Psi_b{}^a - \dot{\bar\Psi}_a{}^b \Psi_b{}^a\right)\right] \\
\!\!&\!\!\!\!&\!\! +\,\frac12\, \sum_{a,b,c} \left[
\frac{(x_a - x_b)^2 }{x_a x_b}\,A_a{}^b A_b{}^a - 2\bar Z^a A_a{}^b Z_b  + 2c\,A_a{}^a+
\frac{x_a + x_b}{x_a x_b x_c}\,A_a{}^b\left(\Psi_b{}^c \bar\Psi_c{}^a + {\bar\Psi}_b{}^c \Psi_c{}^a\right)\right]  \nonumber
\\
\!\!&\!\!\!\!&\!\! -\,\frac14\, \sum_{a,b,c,d}
\frac{1}{x_a x_b x_c x_d}\,\bar\Psi_a{}^b \bar\Psi_b{}^c \Psi_c{}^d \Psi_d{}^a\,. \label{b-Cal2}
\end{eqnarray}

Introduce new variables  $q_a$ and $\Phi_a{}^b$, $\bar\Phi_a{}^b=(\Phi_b{}^a)^*$,
which are defined by expressions
\begin{equation}\label{x-q}
x_a=\exp (q_a) \,,
\end{equation}
\begin{equation}\label{Phi-def}
\Phi_a{}^b:= \frac{\Psi_a{}^b}{\sqrt{x_ax_b}}\,,\qquad
\bar\Phi_a{}^b:= \frac{\bar\Psi_a{}^b}{\sqrt{x_ax_b}}\,.
\end{equation}
In these variables
the kinetic term of the Lagrangian \p{b-Cal2} (first line in \p{b-Cal2}) takes the form
\begin{equation}\label{b-Cal2-k}
\frac12\, \sum_{a,b} \left[ \dot q_a \dot q_a
+ i \left(\bar Z^a \dot Z_a - \dot{\bar Z}{}^a Z_a\right) +
i\left(\bar\Phi_a{}^b \dot\Phi_b{}^a - \dot{\bar\Phi}_a{}^b \Phi_b{}^a\right)\right],
\end{equation}
with flat metric in all sectors.
In this case, the equations of motion of the fields $A_a{}^b$, $a\neq b$ become
\begin{equation}\label{eqA}
A_a{}^b = \frac{1}{4\sinh^2 \Big({\displaystyle\frac{q_a-q_b}{2}}\Big) }\,
\Big[ Z_a \bar Z^b - \cosh \Big({\displaystyle\frac{q_a-q_b}{2}}\Big) \{\Phi, \bar\Phi \}_a{}^b\Big] \qquad\quad \mbox{for } a\neq b\,.
\end{equation}
After eliminating the auxiliary fields $A_a{}^b$, $a\neq b$, the Lagrangian (\ref{b-Cal2}) takes the form
\begin{eqnarray}\nonumber
L^\prime \!\!&\!=\!&\!\!  \frac12\, \sum_{a,b} \left[ \dot q_a \dot q_a
+ i \left(\bar Z^a \dot Z_a - \dot{\bar Z}{}^a Z_a\right) +
i\left(\bar\Phi_a{}^b \dot\Phi_b{}^a - \dot{\bar\Phi}_a{}^b \Phi_b{}^a\right)\right]  \\
\!\!&\!\!\!\!&\!\! -\,\frac18\, \sum_{a\neq b}\,
\frac{\Big[ Z_a \bar Z^b - \cosh \Big({\displaystyle\frac{q_a-q_b}{2}}\Big) \{\Phi, \bar\Phi \}_a{}^b\Big]
\Big[ Z_b \bar Z^a - \cosh \Big({\displaystyle\frac{q_a-q_b}{2}}\Big) \{\Phi, \bar\Phi \}_b{}^a\Big]}{\sinh^2 \Big({\displaystyle\frac{q_a-q_b}{2}}\Big) }\, \nonumber
\\
\!\!&\!\!\!\!&\!\! - \, \frac{1}{4}\,{\rm Tr} \Big( \bar\Phi \bar\Phi \Phi \Phi \Big) \ - \
\sum_{a}\,A_a{}^a\,\Big(Z_a \bar Z^a-\{\Phi, \bar\Phi \}_a{}^a -c \Big)\,. \label{h2-Cal2}
\end{eqnarray}

In the last term of the Lagrangian  (\ref{h2-Cal2}) the fields $A_a{}^a$ are the Lagrange multipliers for $n$ constraints
\begin{equation}\label{eqZ}
Z_a \bar Z^a-\{\Phi, \bar\Phi \}_a{}^a -c=0 \qquad\quad \forall \, a  \qquad (\mbox{no sum over}\ a)\,.
\end{equation}
Due to the constraints \p{eqZ}, the Lagrangian (\ref{h2-Cal2}) possesses
residual invariance under the gauge abelian $[U(1)]^n$ group with the local
parameters $\gamma_a(t)$:
\begin{equation}\label{b-Ab}
Z_a \rightarrow \, \mathrm{e}^{i\gamma_a} Z_a \,, \quad \bar Z^a  \rightarrow \,
\mathrm{e}^{-i\gamma_a}
\bar Z^a\,, \qquad A_a{}^a \rightarrow \, A_a{}^a - \dot \gamma_a \qquad (\mbox{no
sum over}\; a)\,,
\end{equation}
\begin{equation}\label{Psi-Ab}
\Phi_a{}^b \rightarrow \, \mathrm{e}^{i\gamma_a} \Phi_a{}^b \mathrm{e}^{-i\gamma_b}\,, \qquad
\bar\Phi_a{}^b \rightarrow \, \mathrm{e}^{i\gamma_a} \bar\Phi_a{}^b  \mathrm{e}^{-i\gamma_b}\qquad (\mbox{no
sums over}\; a,b)\,.
\end{equation}
Therefore, it is possible to impose a further gauge-fixing condition
\begin{equation}\label{g-Z}
\bar Z^a = Z_a\,.
\end{equation}
In this gauge the system (\ref{h2-Cal2}) is described in the bosonic limit by the action
\begin{equation}\label{el-Cal}
S = \frac12 \int \mathrm{d}t  \Bigg[\, \sum_{a}\dot q_a \dot q_a -
\sum_{a\neq b} \frac{c^2}{4\sinh^2 \Big({\displaystyle\frac{q_a-q_b}{2}}\Big) }\,\Bigg]\,,
\end{equation}
which is just a standard action of the hyperbolic Calogero-Sutherland system of the $A_{n-1}$-root type \cite{C,Su,OP,Per}.

\setcounter{equation}{0}
\section{Hamiltonian formulation}

The Hamiltonian formulation plays an important role in obtaining classical generators of symmetry as
the Noether charges and the subsequent finding of their quantum counterparts.
In this section, we carry out the Hamiltonization of the matrix system with the Lagrangian (\ref{N2Cal-com})
and then, after gauge-fixing at the Hamiltonian level, we find the $\mathcal{N}{=}\, 2$ supersymmetry generators
for the reduced system with the Lagrangian (\ref{h2-Cal2}).

\subsection{Hamiltonian formulation of the matrix system}

The system with the Lagrangian (\ref{N2Cal-com}) is described by the momenta
\begin{equation}\label{P-X}
P_a{}^b = (X^{-1}\nabla X X^{-1})_a{}^b \,,
\end{equation}
\begin{equation}\label{P-Z}
\mathcal{P}^a = \frac{i}2 \, \bar Z^a \,, \quad \bar\mathcal{P}_a = -\frac{i}2 \, Z_a \,,
\qquad
\Pi_a{}^b = \frac{i}2 \,(X^{-1}\bar\Psi X^{-1})_a{}^b \,, \quad
\bar\Pi_a{}^b = \frac{i}2 \,(X^{-1} \Psi X^{-1})_a{}^b \,.
\end{equation}
The momenta of the coordinates $A_a{}^b$ are zero.
The nonvanishing canonical Poisson brackets of the phase space variables are the following:
\begin{equation}\label{PB-X}
\{X_a{}^b, P_c{}^d \}_{\scriptstyle{\mathrm{P}}} =  \delta_a^d \delta_c^b \,,
\end{equation}
\begin{equation}\label{PB-Z}
\{Z_a, \mathcal{P}^b \}_{\scriptstyle{\mathrm{P}}} =  \delta_a^b \,,\quad
\{\bar Z^a, \bar\mathcal{P}_b \}_{\scriptstyle{\mathrm{P}}} =  \delta_b^a \,,
\qquad
\{\Psi_a{}^b, \Pi_c{}^d \}_{\scriptstyle{\mathrm{P}}} =  \delta_a^d \delta_c^b \,, \quad
\{\bar\Psi_a{}^b, \bar\Pi_c{}^d \}_{\scriptstyle{\mathrm{P}}} =  \delta_a^d \delta_c^b \,.
\end{equation}

The canonical Hamiltonian
\begin{equation}\label{t-Ham}
H_{\rm matrix}=\ P_b{}^a \dot X_a{}^b + \mathcal{P}^a \dot Z_a + \bar\mathcal{P}_a \dot{\bar Z}^a +
\Pi_b{}^a \dot\Psi_a{}^b + \bar\Pi_b{}^a\dot{\bar\Psi}_a{}^b - L_{\rm matrix}\ =\ H+ {\rm Tr}\big(A F \big)
\end{equation}
represents the sum of the term
\begin{equation}\label{Ham-matrix}
H = \frac12\,{\rm Tr}\Big(XPXP\Big) + \frac14 \,{\rm Tr}\Big(X^{-1}\bar\Psi X^{-1}\bar\Psi X^{-1}\Psi X^{-1}\Psi\Big)
\end{equation}
and the term ${\rm Tr}\big(A F \big)$ containing the quantities
\begin{equation}\label{F-constr}
F_a{}^b := i[P,X]_a{}^b + Z_a\bar Z^b-\frac12\,\{X^{-1}\Psi, X^{-1}\bar\Psi \}_a{}^b
-\frac12\,\{\Psi X^{-1}, \bar\Psi X^{-1} \}_a{}^b - c\,\delta_a{}^b\,.
\end{equation}
The form of the Hamiltonian  \p{t-Ham} and vanishing momenta of  the coordinates $A_a{}^b$ indicate that
quantities  \p{F-constr} are the constraints
\begin{equation}\label{F-constr1}
F_a{}^b \approx 0
\end{equation}
and the variables $A_a{}^b$ are the Lagrange multipliers for them.

Momenta expressions (\ref{P-Z}) yield additional second class constraints
\begin{equation}\label{const-Z}
G^a := \mathcal{P}^a - \frac{i}2 \, \bar Z^a \approx 0\,, \qquad
\bar G_a := \bar\mathcal{P}_a + \frac{i}2 \, Z_a \approx 0 \,,
\end{equation}
\begin{equation}\label{const-Psi}
\Upsilon_a{}^b := \Pi_a{}^b - \frac{i}2 \,(X^{-1}\bar\Psi X^{-1})_a{}^b\approx 0 \,, \qquad
\bar \Upsilon_a{}^b := \bar\Pi_a{}^b - \frac{i}2 \,(X^{-1} \Psi X^{-1})_a{}^b\approx 0 \,,
\end{equation}
possessing the following nonzero Poisson brackets:
\begin{equation}\label{PB-const-2}
\{ G^a , \bar G_b \}_{\scriptstyle{\mathrm{P}}} =-i\delta^a_b \,,\qquad
\{ \Upsilon_a{}^b , \bar \Upsilon_c{}^d \}_{\scriptstyle{\mathrm{P}}} =-iX^{-1}_{\ \ a}{}^d X^{-1}_{\ \ c}{}^b \,.
\end{equation}
Using the Dirac brackets for the constraints \p{const-Z}, \p{const-Psi}
\begin{eqnarray}\nonumber
\{ A,B\}_{\scriptstyle{\mathrm{D}}} &=&
\{ A,B\}_{\scriptstyle{\mathrm{P}}}-
i \{A  , \Upsilon_a{}^b \}_{\scriptstyle{\mathrm{P}}}X_b{}^c X_d{}^a\{ \bar \Upsilon_c{}^d, B \}_{\scriptstyle{\mathrm{P}}} -i
\{ A , \bar \Upsilon_a{}^b \}_{\scriptstyle{\mathrm{P}}}X_b{}^c X_d{}^a\{ \Upsilon_c{}^d , B \}_{\scriptstyle{\mathrm{P}}} \\ [6pt]
&& \qquad\quad\ +\,i \{A  , G^a \}_{\scriptstyle{\mathrm{P}}} \{ \bar G_a , B \}_{\scriptstyle{\mathrm{P}}} -i \{ A , \bar G_a \}_{\scriptstyle{\mathrm{P}}}\{ G^a , B \}_{\scriptstyle{\mathrm{P}}} \,, \label{DB-const-2}
\end{eqnarray}
we eliminate the momenta $\mathcal{P}^a$, $\bar\mathcal{P}_a$, $\Pi_a{}^b$, $\bar\Pi_a{}^b$.
The nonvanishing  Dirac brackets of residual phase variables take the form
\begin{equation}\label{DB-X}
\{X_a{}^b, P_c{}^d \}_{\scriptstyle{\mathrm{D}}} =  \delta_a^d \delta_c^b \,,
\end{equation}
\begin{equation}\label{DB-P}
\begin{array}{rcl}
\{P_a{}^b, P_c{}^d \}_{\scriptstyle{\mathrm{D}}} &= &  -\frac{i}4\, [X^{-1}(\Psi X^{-1}\bar\Psi + \bar\Psi X^{-1}\Psi)X^{-1}]_a{}^d X^{-1}_{\ \ c}{}^b \\ [5pt]
&& +\, \frac{i}4\, X^{-1}_{\ \ a}{}^d [X^{-1}(\Psi X^{-1}\bar\Psi + \bar\Psi X^{-1}\Psi)X^{-1}]_c{}^b \,,
\end{array}
\end{equation}
\begin{equation}\label{DB-Z}
\{Z_a, \bar Z^b \}_{\scriptstyle{\mathrm{D}}} =  -i\delta_a^b \,,
\qquad
\{\Psi_a{}^b, \bar\Psi_c{}^d \}_{\scriptstyle{\mathrm{D}}} =  -iX_a{}^d X_c{}^b \,,
\end{equation}
\begin{equation}\label{DB-PPs}
\begin{array}{rcl}
\{\Psi_a{}^b, P_c{}^d \}_{\scriptstyle{\mathrm{D}}} &=&  \frac{1}2\, \delta_a^d (X^{-1}\Psi )_c{}^b + \frac{1}2\, \delta_c^b (\Psi X^{-1})_a{}^d\,,\\ [5pt]
\{\bar\Psi_a{}^b, P_c{}^d \}_{\scriptstyle{\mathrm{D}}} &=&  \frac{1}2\, \delta_a^d (X^{-1}\bar\Psi )_c{}^b + \frac{1}2\, \delta_c^b (\bar\Psi X^{-1})_a{}^d\,.
\end{array}
\end{equation}

The residual constraints \p{F-constr1} are ``real''
\begin{equation}\label{real-F}
(F_a{}^b)^* =F_b{}^a
\end{equation}
and form the $u(n)$ algebra with respect to the Dirac brackets \p{DB-const-2}:
\begin{equation}\label{DB-FF}
\{F_a{}^b, F_c{}^d \}_{\scriptstyle{\mathrm{D}}} =-i \delta_a{}^d F_c{}^b + i \delta_c{}^b F_a{}^d \,.
\end{equation}
As a result, the constraints \p{F-constr}, \p{F-constr1} are first class and generate local $\mathrm{U}(n)$ transformations.

\subsection{Hamiltonian formulation of the reduced system}

Local transformations generated by the off-diagonal constraints $F_a{}^b \approx0$, $a{\neq}b$ in the set  \p{F-constr}, \p{F-constr1}
are fixed by imposing the gauges \p{X-fix}, \p{X-fix-com}.
After using the expansions
\begin{equation}\label{XP-exp}
X_a{}^b =x_a \delta_a{}^b + x_a{}^b\,,
\qquad
P_a{}^b = \mathrm{p}_a \delta_a{}^b + \mathrm{p}_a{}^b\,,
\end{equation}
where $x_a{}^b$ and $\mathrm{p}_a{}^b$ represent the off-diagonal matrix terms, i.e.  $x_a{}^a=\mathrm{p}_a{}^a=0$ at fixed index $a$,
the gauge fixing  \p{X-fix}, \p{X-fix-com} takes the form
\begin{equation}\label{x-fix}
x_a{}^b\approx 0\,.
\end{equation}
When the gauge fixing conditions \p{x-fix} are fulfilled,
the constraints $F_a{}^b \approx0$, $a{\neq}b$ allow one to express the momenta $\mathrm{p}_a{}^b$ through the remaining  phase variables:
\begin{equation}\label{p-exp}
\mathrm{p}_a{}^b= -\frac{i\,Z_a\bar Z^b}{x_a-x_b}+\frac{i\,(x_a+x_b)\,\{\Phi,\bar\Phi\}_a{}^b}{2(x_a-x_b)\sqrt{x_ax_b}}\,,
\end{equation}
where we use the odd matrix variables \p{Phi-def}.
Thus, introducing the Dirac brackets, we can eliminate the variables
$x_a{}^b$, $\mathrm{p}_a{}^b$ by means of expressions \p{x-fix}, \p{p-exp}.
Moreover, these Dirac brackets coincide with the Dirac brackets (\ref{DB-X})-(\ref{DB-PPs})
since the gauge fixing conditions \p{x-fix} include only $x_a{}^b$.

As a result, after eliminating the variables $x_a{}^b$, $\mathrm{p}_a{}^b$, the considered system is described
by $n$ even real variables $x_a$, $\mathrm{p}_a$, $n^2$ odd complex variables $\Phi_a{}^b$ and $n$ even complex variables $Z_a$.
Their nonvanishing Dirac brackets are
\begin{eqnarray}\label{DB-xp}
\{x_a, \mathrm{p}_b \}^{'}_{\scriptstyle{\mathrm{D}}} &=&  \delta_{ab} \,,
\\ [6pt]
\label{DB-Z1}
\{Z_a, \bar Z^b \}^{'}_{\scriptstyle{\mathrm{D}}} &=&  -i\,\delta_a^b \,,
\\ [6pt]
\label{DB-Ph}
\{\Phi_a{}^b, \bar\Phi_c{}^d \}^{'}_{\scriptstyle{\mathrm{D}}} &=&  -i\,\delta_a^d \delta_c^b \,.
\end{eqnarray}
Point out that in the gauge \p{x-fix} the momenta $\mathrm{p}_a$ commute with each other (compare with \p{DB-P}).
Besides, the Grassmannian quantities $\Phi_a{}^b$ commute with $\mathrm{p}_a$ (compare with \p{DB-PPs}).
Also, their Dirac brackets \p{DB-Ph} are proportional to the Kronecker symbols, in contrast to the Dirac bracket for $\Psi_a{}^b$ (see \p{DB-Z}).

Let us introduce the variables $q_a$ by relations \p{x-q}. The canonical momenta $p_{\,a}$ of the coordinates  $q_a$ are defined by
\begin{equation}\label{p-q}
p_{\,a}=x_a \mathrm{p}_a \,,\qquad \{q_a,p_{\,b}\}^{'}_{\scriptstyle{\mathrm{D}}} =\delta_{ab}\,.
\end{equation}
In the variables $q_a$, $p_{\,a}$ and \p{Phi-def} and after the gauge-fixing \p{x-fix}, \p{p-exp}
the Hamiltonian \p{Ham-matrix} takes the form
\begin{equation}\label{Ham-fix}
\mathrm{H} = \frac12\,\sum_{a}p_a p_a +
\frac18\,\sum_{a\neq b} \frac{R_a{}^bR_b{}^a}{\sinh^2 \Big({\displaystyle\frac{q_a-q_b}{2}}\Big)} +
\frac14 \,{\rm Tr}\Big(\bar\Phi \bar\Phi \Phi \Phi\Big)\,,
\end{equation}
where
\begin{equation}\label{T-def}
R_a{}^b := Z_a\bar Z^b- \cosh\left(\frac{q_a-q_b}{2}\right)\{ \Phi, \bar\Phi \}_a{}^b\,.
\end{equation}
The residual first class constraints in the set \p{F-constr}, \p{F-constr1} are $n$ diagonal constraints
\begin{equation}\label{F-constr-d}
F_a := F_a{}^a =R_a{}^a -c= Z_a\bar Z^a- \{ \Phi, \bar\Phi \}_a{}^a - c\approx 0\qquad\mbox{(no summation over $a$)}\,,
\end{equation}
which form an abelian algebra with respect to the Dirac brackets \p{DB-Ph}
\begin{equation}
\label{DB-constr1}
\{F_a , F_b \}^{'}_{\scriptstyle{\mathrm{D}}} =  0
\end{equation}
and generate the $[\mathrm{U}(1)]^n$ gauge transformations of $Z_a$ and $\Phi_a{}^b$.

Similarly to \p{XP-exp} we can use the expansions of the Grassmannian matrix quantities \p{Phi-def}
on the diagonal and off-diagonal parts:
\begin{equation}\label{Phi-exp}
\Phi_a{}^b =\varphi_a \delta_a{}^b + \phi_a{}^b\,,
\qquad
\bar\Phi_a{}^b =\bar\varphi_a \delta_a{}^b + \bar\phi_a{}^b\,,
\end{equation}
where $\phi_a{}^a=\bar\phi_a{}^a=0$ at fixed index $a$.
The Dirac brackets \p{DB-Ph} of the diagonal quantities $\varphi_a$, $\bar\varphi_a$ and
the off-diagonal ones $\phi_a{}^b$, $\bar\phi_a{}^b$ have the form
\begin{equation}
\label{DB-Ph1}
\{\varphi_a , \bar\varphi_b \}^{'}_{\scriptstyle{\mathrm{D}}} =  -i\,\delta_{a b}\,,\qquad
\{\phi_a{}^b, \bar\phi_c{}^d \}^{'}_{\scriptstyle{\mathrm{D}}} =  -i\,\delta_a^d \delta_c^b \,.
\end{equation}
The constraints \p{F-constr-d} involve only the off-diagonal fermions $\phi$, $\bar\phi$:
\begin{equation}\label{F-constr-d1}
F_a = Z_a\bar Z^a- \{ \phi, \bar\phi \}_a{}^a - c\approx 0\qquad\mbox{(no summation over $a$)}\,.
\end{equation}
In the variables $\varphi$, $\bar\varphi$, $\phi$, $\bar\phi$ the Hamiltonian \p{Ham-fix} takes the form
\begin{eqnarray}
\mathrm{H} &=& \frac12\,\sum_{a}p_a p_a +
\frac18\,\sum_{a\neq b} \frac{Z_a \bar Z^a Z_b \bar Z^b}{\sinh^2 \Big({\displaystyle\frac{q_a-q_b}{2}}\Big)}
\nonumber\\
&&
+\,
\frac14\,\sum_{a\neq b} \frac{\coth \Big({\displaystyle\frac{q_a-q_b}{2}}\Big) }
{\sinh \Big({\displaystyle\frac{q_a-q_b}{2}}\Big)}\,Z_a \bar Z^b
\Big[(\varphi_a-\varphi_b)\bar\phi_b{}^a+ (\bar\varphi_a-\bar\varphi_b)\phi_b{}^a - \{\phi,\bar\phi \}_b{}^a\Big]
\nonumber\\
&&
-\,
\frac18\,\sum_{a\neq b}
\frac{1}
{\sinh^2 \Big({\displaystyle\frac{q_a-q_b}{2}}\Big)}\,
\Big[2(\varphi_a-\varphi_b)(\bar\varphi_a-\bar\varphi_b)\phi_a{}^b\bar\phi_b{}^a
-2(\varphi_a-\varphi_b)\bar\phi_a{}^b \{\phi,\bar\phi \}_b{}^a
\nonumber \\
&&\qquad\qquad\qquad \qquad\qquad\quad
-2(\bar\varphi_a-\bar\varphi_b)\phi_a{}^b \{\phi,\bar\phi \}_b{}^a
- \{\phi,\bar\phi \}_a{}^b \{\phi,\bar\phi \}_b{}^a \Big]
\nonumber\\
&&
-\,
\frac18\,\sum_{a} \{\phi,\bar\phi \}_a{}^a \{\phi,\bar\phi \}_a{}^a\,.
\label{Ham-fix1}
\end{eqnarray}
We see that in the bosonic limit and on the shell of constraints \p{F-constr-d1}
the Hamiltonian \p{Ham-fix1} is the Hamiltonian of the model \p{el-Cal}
which describes the hyperbolic Calogero-Sutherland $A_{n-1}$-root system \cite{C,Su,OP,Per}.

\setcounter{equation}{0}
\section{Classical generators of the ${\mathcal N}{=}\,2$ supersymmetry }

Since the system with the Lagrangian (\ref{N2Cal-com}) considered here was obtained from the ${\mathcal N}{=}\,2$ superfield action \cite{FIL19},
it possesses the ${\mathcal N}{=}\,2$ supersymmetry invariance.
Supersymmetry transformations of the matrix component fields are
\footnote{These transformations are the sum of usual supersymmetry transformations
of the component fields and compensating gauge transformation for preservation of the Wess-Zumino gauge.}
\begin{equation}\label{tr-susy}
\begin{array}{rcl}
\delta X &=& \varepsilon \Psi -  \bar\varepsilon \bar\Psi\,, \\ [6pt]
\delta \Psi &=& \bar\varepsilon\left(-i\nabla X + {\displaystyle \frac12}\,\Psi X^{-1}\bar\Psi
-{\displaystyle \frac12}\,\bar\Psi X^{-1}\Psi\right),\\ [6pt]
\delta \bar\Psi &=& \varepsilon\left(i\nabla X + {\displaystyle \frac12}\,\Psi X^{-1}\bar\Psi
-{\displaystyle \frac12}\,\bar\Psi X^{-1}\Psi\right), \\ [6pt]
\delta Z&=&0\,,\quad \delta \bar Z\ =\ 0\,,\qquad \delta A\ =\ 0\,,
\end{array}
\end{equation}
where $\varepsilon$, $\bar\varepsilon=(\varepsilon)^*$ is the complex Grassmannian parameter.
The corresponding Noether charges have the form
\begin{equation}\label{Q-matrix}
Q = {\rm Tr}\big( P\Psi\big)\,,\qquad \bar Q = {\rm Tr}\big( P\bar\Psi\big)\,,
\end{equation}
where the matrix momentum $P_a{}^b$ is defined in \p{P-X}.
The supercharges \p{Q-matrix} and the Hamiltonian $H$ presented in \p{Ham-matrix}
form the ${\mathcal N}{=}\,2$ $d{=}\,1$ superalgebra with respect to the Dirac brackets \p{DB-X}-\p{DB-PPs}:
\begin{equation}\label{N2-susy-matrix}
\{ Q, \bar Q \}_{\scriptstyle{\mathrm{D}}} = -2i\,H\,,\qquad \{ Q, H \}_{\scriptstyle{\mathrm{D}}}=\{ \bar Q, H \}_{\scriptstyle{\mathrm{D}}}=0\,.
\end{equation}

Now we find the supersymmetry generators for the reduced system,
which is described by the Hamiltonian \p{Ham-fix} and the first class constraints \p{F-constr-d}.
Inserting the gauge fixing conditions \p{x-fix}, \p{p-exp} in \p{Q-matrix} and using the variables \p{Phi-def}, \p{p-q}, we obtain
\begin{equation}\label{Q}
\mathrm{Q} =\sum\limits_{a} p_a \Phi_a{}^a-\frac{i}{2}\sum\limits_{a\neq b}
\frac{ R_a{}^b \Phi_b{}^a}{\sinh\Big({\displaystyle\frac{q_a-q_b}{2}}\Big)} \,,
\qquad
\bar \mathrm{Q} =\sum\limits_{a} p_a \bar\Phi_a{}^a-\frac{i}{2}\sum\limits_{a\neq b}
\frac{ R_a{}^b \bar\Phi_b{}^a}{\sinh\Big({\displaystyle\frac{q_a-q_b}{2}}\Big)} \,.
\end{equation}
In the Grassmannian variables $\varphi$, $\bar\varphi$, $\phi$, $\bar\phi$, defined in \p{Phi-exp}, these generators take the form
\begin{eqnarray}\nonumber
\mathrm{Q} &= &\sum\limits_{a} p_a \varphi_a-\frac{i}{2}\sum\limits_{a\neq b}
\frac{ Z_a\bar Z^b \phi_b{}^a}{\sinh\Big({\displaystyle\frac{q_a-q_b}{2}}\Big)} \\
&& \qquad \qquad \qquad +\,\frac{i}{2}\sum\limits_{a\neq b}
\coth \Big({\displaystyle\frac{q_a-q_b}{2}}\Big)
\Big[ (\varphi_a-\varphi_b)\bar\phi_a{}^b +\{\phi,\bar\phi\}_a{}^b
\Big]\,\phi_b{}^a\,,
\label{Q1}\\
\nonumber
\bar \mathrm{Q} &= &\sum\limits_{a} p_a \bar\varphi_a-\frac{i}{2}\sum\limits_{a\neq b}
\frac{ Z_a\bar Z^b \bar\phi_b{}^a}{\sinh\Big({\displaystyle\frac{q_a-q_b}{2}}\Big)} \\
&& \qquad \qquad \qquad +\,\frac{i}{2}\sum\limits_{a\neq b}
\coth\Big({\displaystyle\frac{q_a-q_b}{2}}\Big)
\Big[ (\bar\varphi_a-\bar\varphi_b)\phi_a{}^b  +\{\phi,\bar\phi\}_a{}^b
\Big]\,\bar\phi_b{}^a\,,
\label{bQ1}
\end{eqnarray}
Using the Dirac brackets \p{p-q}, \p{DB-Z1}, \p{DB-Ph1}, we find that the supercharges $\mathrm{Q}$, $\bar \mathrm{Q}$
form the superalgebra
\footnote{
In obtaining  \p{DB-QQ}-\p{DB-HbQ} the equality
$$
\coth (y_a-y_b )\coth (y_b-y_c ) \ + \
\coth (y_b-y_c ) \coth (y_c-y_a )
\ +\ \coth (y_c-y_a )\coth (y_a-y_b ) = -1
$$
was used, where $y_a\neq y_b\neq y_c\neq y_a$.
}
\begin{eqnarray}\label{DB-QQ}
\{\mathrm{Q}, \mathrm{Q} \}^{'}_{\scriptstyle{\mathrm{D}}} &=&  -\frac{i}{4}\sum\limits_{a\neq b}
\frac{ \phi_a{}^b \phi_b{}^a}{\sinh^2\Big({\displaystyle\frac{q_a-q_b}{2}}\Big)}\,\Big( F_a-F_b\Big) \,,
\\ [6pt]
\label{DB-bQbQ}
\{\bar \mathrm{Q} , \bar \mathrm{Q} \}^{'}_{\scriptstyle{\mathrm{D}}} &=&  -\frac{i}{4}\sum\limits_{a\neq b}
\frac{ \bar\phi_a{}^b \bar\phi_b{}^a}{\sinh^2\Big({\displaystyle\frac{q_a-q_b}{2}}\Big)}\,\Big( F_a-F_b\Big) \,,
\\ [6pt]
\label{DB-QbQ}
\{ \mathrm{Q} , \bar \mathrm{Q} \}^{'}_{\scriptstyle{\mathrm{D}}} &=&  -2i\,\mathrm{H}  -\frac{i}{4}\sum\limits_{a\neq b}
\frac{ \phi_a{}^b \bar\phi_b{}^a}{\sinh^2\Big({\displaystyle\frac{q_a-q_b}{2}}\Big)}\,\Big( F_a-F_b\Big) \,,
\\ [6pt]
\label{DB-HQ}
\{ \mathrm{Q}, \mathrm{H} \}^{'}_{\scriptstyle{\mathrm{D}}} &=&
-\frac{1}{8}\sum\limits_{a\neq b}
\frac{ R_a{}^b \phi_b{}^a}{\sinh^3\Big({\displaystyle\frac{q_a-q_b}{2}}\Big)}\,\Big( F_a-F_b\Big) \,,
\\ [6pt]
\label{DB-HbQ}
\{ \bar\mathrm{Q}, \mathrm{H} \}^{'}_{\scriptstyle{\mathrm{D}}} &=&
-\frac{1}{8}\sum\limits_{a\neq b}
\frac{ R_a{}^b \bar\phi_b{}^a}{\sinh^3\Big({\displaystyle\frac{q_a-q_b}{2}}\Big)}\,\Big( F_a-F_b\Big) \,,
\end{eqnarray}
where the Hamiltonian $\mathrm{H}$ is defined in \p{Ham-fix1} and the constraints $F_a\approx0$ are given in \p{F-constr-d1}.
Thus, the quantities  $\mathrm{H}$, $\mathrm{Q}$, $\bar \mathrm{Q}$,
defined in \p{Ham-fix}, \p{Q} (or the same \p{Ham-fix1}, \p{Q1}, \p{bQ1}),
form the $\mathcal{N}{=}\,2$ superalgebra with respect to the Dirac brackets \p{p-q}, \p{DB-Z1}, \p{DB-Ph1}
on the shell of the first class constraints \p{F-constr-d1}.
Moreover, the generators $\mathrm{H}$, $\mathrm{Q}$, $\bar \mathrm{Q}$
are gauge invariant: they have the vanishing Dirac brackets with the first class constraints  \p{F-constr-d1},
\begin{equation}
\label{DB-constr1-Q}
\{\mathrm{Q} , F_a \}^{'}_{\scriptstyle{\mathrm{D}}} =  \{\bar\mathrm{Q} , F_a \}^{'}_{\scriptstyle{\mathrm{D}}} =
\{ \mathrm{H} , F_a \}^{'}_{\scriptstyle{\mathrm{D}}} =0 \,.
\end{equation}

It should be noted that the terms in the supercharges \p{Q1}, \p{bQ1} describing the interaction of particles
(i.e. \p{Q1}, \p{bQ1} without the first terms $\sum p_a \varphi_a$, $\sum p_a \bar\varphi_a$)
are zero when the off-diagonal matrix fermions $\phi_a{}^b$, $\bar\phi_a{}^b$ vanish.
In Section~6, it will be shown that in the quantum case the situation is different:
in the quantum supercharges there are terms with interaction that depend only on diagonal matrix variables
and generate the $\mathcal{N}{=}\,2$ superalgebra in this subsector.

Similar to the Lagrangian consideration in Section~2, we can make the gauge-fixing \p{g-Z} for the first class constraints \p{F-constr-d}
(or \p{F-constr-d1}).
Then, the components of the spinor $Z_a$ become real and are expressed  through the remaining variables by the following expressions:
\begin{equation}\label{Z-sqrt}
Z_a= \sqrt{c+\{ \phi, \bar\phi \}_a{}^a}   \qquad\mbox{(no summation over $a$)}\,.
\end{equation}
In this gauge the supercharges \p{Q1}, \p{bQ1} take the form
\begin{eqnarray}\nonumber
\mathrm{Q} &= &\sum\limits_{a} p_a \varphi_a-\frac{i}{2}\sum\limits_{a\neq b}
\frac{ \sqrt{c+\{ \phi, \bar\phi \}_a{}^a}\,\sqrt{c+\{ \phi, \bar\phi \}_b{}^b}\, \phi_b{}^a}{\sinh\Big({\displaystyle\frac{q_a-q_b}{2}}\Big)} \\
&& \qquad \qquad \qquad +\,\frac{i}{2}\sum\limits_{a\neq b}
\coth \Big({\displaystyle\frac{q_a-q_b}{2}}\Big)
\Big[ (\varphi_a-\varphi_b)\bar\phi_a{}^b +\{\phi,\bar\phi\}_a{}^b
\Big]\,\phi_b{}^a\,,
\label{Q2}\\
\nonumber
\bar \mathrm{Q} &= &\sum\limits_{a} p_a \bar\varphi_a-\frac{i}{2}\sum\limits_{a\neq b}
\frac{ \sqrt{c+\{ \phi, \bar\phi \}_a{}^a}\,\sqrt{c+\{ \phi, \bar\phi \}_b{}^b}\,
\bar\phi_b{}^a}{\sinh\Big({\displaystyle\frac{q_a-q_b}{2}}\Big)} \\
&& \qquad \qquad \qquad +\,\frac{i}{2}\sum\limits_{a\neq b}
\coth\Big({\displaystyle\frac{q_a-q_b}{2}}\Big)
\Big[ (\bar\varphi_a-\bar\varphi_b)\phi_a{}^b  +\{\phi,\bar\phi\}_a{}^b
\Big]\,\bar\phi_b{}^a\,.
\label{bQ2}
\end{eqnarray}
The supercharges \p{Q2}, \p{bQ2}  have a complicated structure:
they contain degrees higher than third with respect to the Grassmannian quantities $\phi_a{}^b$, $\bar\phi_a{}^b$
due to the presence of the square roots in the second terms in them.
Following \cite{KLS-18b} we can introduce new variables
\begin{equation}\label{xi-def}
\xi_a{}^b= \phi_a{}^b\sqrt{\frac{c+\{ \phi, \bar\phi \}_b{}^b}{c+\{ \phi, \bar\phi \}_a{}^a}}\,,  \qquad
\bar\xi_a{}^b= \bar\phi_a{}^b\sqrt{\frac{c+\{ \phi, \bar\phi \}_b{}^b}{c+\{ \phi, \bar\phi \}_a{}^a}}\,,
\end{equation}
in which the supercharges \p{Q1}, \p{bQ1} are of no higher than the third degree with respect to all Grassmann variables:
\begin{eqnarray}\nonumber
\mathrm{Q} &= &\sum\limits_{a} p_a \varphi_a-\frac{i}{2}\sum\limits_{a\neq b}
\frac{ \left(c+\{ \xi, \bar\xi \}_b{}^b\right) \xi_b{}^a}{\sinh\Big({\displaystyle\frac{q_a-q_b}{2}}\Big)} \\
&& \qquad \qquad \qquad +\,\frac{i}{2}\sum\limits_{a\neq b}
\coth \Big({\displaystyle\frac{q_a-q_b}{2}}\Big)
\Big[ (\varphi_a-\varphi_b)\bar\xi_a{}^b +\{\xi,\bar\xi\}_a{}^b
\Big]\,\xi_b{}^a\,,
\label{Q3}\\
\nonumber
\bar \mathrm{Q} &= &\sum\limits_{a} p_a \bar\varphi_a-\frac{i}{2}\sum\limits_{a\neq b}
\frac{ \left(c+\{ \xi, \bar\xi \}_b{}^b\right)
\bar\xi_b{}^a}{\sinh\Big({\displaystyle\frac{q_a-q_b}{2}}\Big)} \\
&& \qquad \qquad \qquad +\,\frac{i}{2}\sum\limits_{a\neq b}
\coth\Big({\displaystyle\frac{q_a-q_b}{2}}\Big)
\Big[ (\bar\varphi_a-\bar\varphi_b)\xi_a{}^b  +\{\xi,\bar\xi\}_a{}^b
\Big]\,\bar\xi_b{}^a\,.
\label{bQ3}
\end{eqnarray}
The supercharges \p{Q3}, \p{bQ3} coincide exactly with the ${\mathcal N}{=}\,2$ supersymmetry generators presented in \cite{Pr-2019}, \cite{Kr-19}.
Despite the absence of irrationalities in the supercharges \p{Q3}, \p{bQ3} and the absence of additional gauge symmetries in the system,
they have a very unusual property:
the Grassmann quantities \p{xi-def} do not turn into each other under conjugation,
\begin{equation}\label{xi-conj}
(\xi_a{}^b)^*= \bar\xi_b{}^a\,\frac{c+\{ \phi, \bar\phi \}_b{}^b}{c+\{ \phi, \bar\phi \}_a{}^a}\,.
\end{equation}
In our opinion, this fact is a serious obstacle to the construction of its quantum realization and
the corresponding quantum counterparts of the supercharges.
For this reason, we will work with the supercharges \p{Q} (or \p{Q1}, \p{bQ1}) below.

\setcounter{equation}{0}
\section{Lax pair and conserved charges}

Classical dynamics of the reduced system with
the Hamiltonian $\mathrm{H}$ and the generators of the ${\mathcal N}{=}\,2$ supertranslations
$\mathrm{Q}$, $\bar\mathrm{Q}$, defined in \p{Ham-fix1}, \p{Q1}, \p{bQ1},
can be represented in the Lax representation \cite{Lax}.
To do this, it is necessary to consider the following pair of the $n{\times}n$ matrices:
\begin{equation}
\label{L-matr}
L_{a}{}^{b} \ = \ p_a\, \delta_{a}{}^{b} \ - \ i\left( 1- \delta_{a}^{b} \right)
\frac{ R_a{}^b }{2\sinh\Big({\displaystyle\frac{q_a-q_b}{2}}\Big)}\,,
\end{equation}
\begin{equation}
\label{M-matr}
M_{a}{}^{b} \ = \
-\frac{1}{4}\,\{\Phi,\bar\Phi\}_{a}{}^{a}\delta_{a}{}^{b} \ - \ \frac{1}{4}\left( 1- \delta_{a}^{b} \right)
\left(\frac{ \cosh\Big({\displaystyle\frac{q_a-q_b}{2}}\Big)}
{\sinh^2\Big({\displaystyle\frac{q_a-q_b}{2}}\Big)}\,R_a{}^b +\{\Phi,\bar\Phi\}_{a}{}^{b}\right).
\end{equation}
Then, the evolution of the matrix $L$,
\begin{equation}
\label{L-der-eq}
\dot L_{a}{}^{b} \ = \ \{ L_{a}{}^{b}, \mathrm{H} \}^{'}_{\scriptstyle{\mathrm{D}}} \,,
\end{equation}
is represented by the matrix commutator
\begin{equation}
\label{L-eq}
\dot L_{a}{}^{b} \ = \ -i [L,M]_{a}{}^{b}
-i\left( 1- \delta_{a}^{b} \right)\frac{ L_a{}^b\left(F_a-F_b\right)}{4\sinh^2\Big({\displaystyle\frac{q_a-q_b}{2}}\Big)} \,,
\end{equation}
where  $F_a$ are the constraints defined in  \p{F-constr-d1}.
Thus, on the shell of constraints \p{F-constr-d1} $F_a\approx 0$ these matrices $L_{a}{}^{b}$ and $M_{a}{}^{b}$
form the Lax pair.

Similar to equations \p{L-der-eq}, \p{L-eq}, the equations of motion
of the fermionic matrix variables $\Phi_{a}{}^{b}$, $\bar\Phi_{a}{}^{b}$
are also represented as commutators of these matrices and the matrix $M$ \p{M-matr}:
\begin{equation}
\label{Ps-eq}
\dot \Phi_{a}{}^{b} \ = \ \{ \Phi_{a}{}^{b}, \mathrm{H} \}^{'}_{\scriptstyle{\mathrm{D}}} \ = \ -i [\Phi,M]_{a}{}^{b} \,,\qquad
\dot {\bar\Phi}_{a}{}^{b} \ = \ \{ {\bar\Phi}_{a}{}^{b}, \mathrm{H} \}^{'}_{\scriptstyle{\mathrm{D}}} \ = \ -i [{\bar\Phi},M]_{a}{}^{b} \,.
\end{equation}

After obtaining the Lax representations for the equations of motion \p{L-eq}, \p{Ps-eq}, the conserved charges are easily found.
So equations \p{L-eq} imply that the trace of the $\mathrm{k}$-th degree of the matrix \p{L-matr}
\begin{equation}
\label{Ik-def}
I_\mathrm{k} := {\mathrm{Tr}} (L^\mathrm{k})\,,\qquad \mathrm{k}=1,\ldots,n
\end{equation}
satisfies the equation
\begin{equation}
\label{Ik-eq}
\dot I_\mathrm{k} = \frac{i\mathrm{k}}{4}\sum\limits_{a\neq b}
\frac{ (L^{\mathrm{k}})_a{}^b }{\sinh^2\Big({\displaystyle\frac{q_a-q_b}{2}}\Big)}\,\Big( F_a-F_b\Big)\,.
\end{equation}
Thus, on the shell of constraints \p{F-constr-d1}, the quantities \p{Ik-def} are conserved:
\begin{equation}
\label{Ik-const}
\dot I_\mathrm{k} \approx 0\,.
\end{equation}
Moreover, equations \p{Ps-eq} show that odd quantities
\begin{equation}
\label{Ik-f-def}
\mathcal{I}_{\,\mathrm{k}} := {\mathrm{Tr}} (\Phi L^\mathrm{k})\,, \qquad \bar{\mathcal{I}}_{\,\mathrm{k}} := {\mathrm{Tr}} (\bar\Phi L^\mathrm{k})
\,,\qquad \mathrm{k}=0,1,\ldots,n-1
\end{equation}
are also conserved:
\begin{equation}
\label{Ik-f-const}
\dot {\mathcal{I}}_{\,\mathrm{k}} = 0\,,\qquad \dot {\bar{\mathcal{I}}}_{\,\mathbf{k}} = 0\,.
\end{equation}

Equations \p{L-eq}, \p{Ps-eq} lead to a stronger consequence: the trace
\begin{equation}
\label{J-f-def}
\mathcal{J} := {\mathrm{Tr}} (\mathcal{F})
\end{equation}
of any polynomial function
$\mathcal{F}(L,\Phi,\bar\Phi)$ of the matrix variables $L_{a}{}^{b}$, ${\Phi}_{a}{}^{b}$, ${\bar\Phi}_{a}{}^{b}$
is a conserved quantity on the shell of constraints \p{F-constr-d1}:
\begin{equation}
\label{F-conser}
\dot \mathcal{J} \approx 0\,.
\end{equation}
In particular, the third term in the Hamiltonian \p{Ham-fix}
\begin{equation}\label{Ham-fix3}
J := -\frac18 \,{\rm Tr}\Big(\{\Phi,\bar\Phi\}\{\Phi,\bar\Phi\}\Big)
\end{equation}
is conserved, $\dot J=0$. The total Hamiltonian \p{Ham-fix} is the sum of this conserved charge and half of the second member in the set \p{Ik-def}:
\begin{equation}\label{Ham-fix-c}
\mathrm{H} = \frac12\,I_2 +J\,.
\end{equation}
In addition, the supercharges \p{Q} are exactly the first members in the sets \p{Ik-f-def}:
\begin{equation}\label{Q-c}
\mathrm{Q} =\mathcal{I}_{\,1}\,,
\qquad \bar \mathrm{Q} =\bar{\mathcal{I}}_{\,1} \,.
\end{equation}

Deriving the Lax pair and finding the set of conserved charges \p{J-f-def} paves the way for analyzing the integrability 
of the $\mathcal{N}{=}\,2$ supersymmetric system considered here.
To do this, it is useful to apply the particle scattering property for the Calogero-Sutherland hyperbolic system
that is the bosonic core of the model studied here.
Namely, with the initial ordering $q_a\,{>}\,q_b$ at $a\,{>}\,b$ in the Calogero-Sutherland hyperbolic system, in the limit of infinite time $t\,{\to}\,\infty$ all interparticle distances run to infinity: $(q_a-g_b)\,{\to}\,\infty$ \cite{CalRM,Per}.
In this limit, matrix \p{L-matr} takes a simple form
\begin{equation}
\label{L-matr-lim}
L_{a}{}^{b} \ = \ p_a\, \delta_{a}{}^{b} \ + \ \frac{i}{2}\,
{\rm sgn}(a-b) \{\Phi,\bar\Phi\}_a{}^b \,.
\end{equation}
The explicit form of this matrix tells us that 
when analyzing the integrability of the considered system, in addition to $n$ even conserved charges 
we can take, for example, $n^2$ complex odd charges
\begin{equation}
\label{Ik-f2-def}
\mathcal{I}_{\,\mathrm{k},\mathrm{j}} = {\mathrm{Tr}} \Big(\Phi \{\Phi,\bar\Phi\}^{\mathrm{k}}L^\mathrm{j}\Big)\,, \qquad 
\mathrm{k},\mathrm{j}=0,1,\ldots,n-1
\end{equation}
and $n^2$ their complex conjugated ones ($\mathcal{I}_{\,\mathrm{k}}=\mathcal{I}_{\,0,\mathrm{k}}$ in \p{Ik-f-def}).
The analysis of the superalgebra of the conserved charges and the integrability of 
the considered many-particle supersymmetric system will be the subject of the next article.

It should be noted that the structure of the conserved charges in the considered supersymmetric system \p{F-conser},
in particular, the supercharges \p{Q-c} as ${\mathrm{Tr}} (\Phi L)$,
is similar to the form of the charges in the trigonometric (non-matrix) supersymmetric system studied in \cite{DeLaMa}.

\setcounter{equation}{0}
\section{Quantum generators of the  $\mathcal{N}{=}\,2$ superalgebra}

Quantum $\mathcal{N}{=}\,2$ supersymmetric hyperbolic Calogero-Sutherland system is described by the operators
$$
\mathbf{q}_a\,, \ \mathbf{p}_{\,a}\,, \qquad\mathbf{Z}_a\,, \ \bar {\mathbf{Z}}^a\,, \qquad\bm{\Phi}_a{}^b\,, \ \bar{\bm{\Phi}}_a{}^b\,,
$$
which are quantum operators corresponding to the classical phase space variables
${q}_a$, ${p}_{\,a}$, ${Z}_a$, $\bar {{Z}}^a$, ${\Phi}_a{}^b$, $\bar{{\Phi}}_a{}^b$.
The Dirac brackets \p{p-q}, \p{DB-Z1}, \p{DB-Ph} produce the following canonical
(anti)commutation relations:
\begin{equation}
\label{al-qu}
[\mathbf{q}_a,\mathbf{p}_{\,b}] =i\,\delta_{ab} \,,
\qquad
[\mathbf{Z}_a, \bar {\mathbf{Z}}^b ] =  \delta_a^b \,,
\qquad
\{ \bm{\Phi}_a{}^b, \bar{\bm{\Phi}}_c{}^d \} =  \delta_a^d \delta_c^b \,.
\end{equation}
Similar to \p{Phi-exp} we can use the expansions for the operators
\begin{equation}\label{Phi-exp-op}
\bm{\Phi}_a{}^b ={\bm{\varphi}}_a \delta_a{}^b + \bm{\phi}_a{}^b\,,
\qquad
\bar{\bm{\Phi}}_a{}^b =\bar{\bm{\varphi}}_a \delta_a{}^b + \bar{\bm{\phi}}_a{}^b\,,
\end{equation}
where ${\bm{\phi}}_a{}^a=\bar{\bm{\phi}}_a{}^a=0$ at fixed index $a$.
Then, the  anticommutators in \p{al-qu} take the form
\begin{equation}
\label{al-Ph1}
\{{\bm{\varphi}}_a , \bar{\bm{\varphi}}_b \} =  \delta_{a b}\,,\qquad
\{{\bm{\phi}}_a{}^b, \bar{\bm{\phi}}_c{}^d \} =  \delta_a^d \delta_c^b \,.
\end{equation}
Below we will use the coordinate representation for the operators $\mathbf{q}_a$, $\mathbf{p}_{\,a}$:
\begin{equation}
\label{coord-repr}
\mathbf{q}_a ={q}_a\,,
\qquad
\mathbf{p}_{\,a} = -i \partial/\partial {q}_a\,.
\end{equation}

Taking the Weyl-ordering in the quantum counterpart of the classical supercharges \p{Q}, we obtain the quantum supercharges:
\begin{eqnarray}\label{Q-q}
\mathbf{Q} &=&\sum\limits_{a} \mathbf{p}_a {\bm{\varphi}}_a -\frac{i}{4}\sum\limits_{a\neq b}
\coth\Big({\displaystyle\frac{q_a-q_b}{2}}\Big)\Big({\bm{\varphi}}_a - {\bm{\varphi}}_b\Big)
-\frac{i}{2}\sum\limits_{a\neq b}
\frac{ \mathbf{R}_a{}^b {\bm{\phi}}_b{}^a}{\sinh\Big({\displaystyle\frac{q_a-q_b}{2}}\Big)} \,,
\\
\label{bQ-q}
\bar{\mathbf{Q}} &=&\sum\limits_{a} \mathbf{p}_a \bar{\bm{\varphi}}_a  + \frac{i}{4}\sum\limits_{a\neq b}
\coth\Big({\displaystyle\frac{q_a-q_b}{2}}\Big)\Big(\bar{\bm{\varphi}}_a  - \bar{\bm{\varphi}}_b \Big)  +\frac{i}{2}\sum\limits_{a\neq b}
\frac{ \bar{\bm{\phi}}_a{}^b \mathbf{R}_b{}^a }{\sinh\Big({\displaystyle\frac{q_a-q_b}{2}}\Big)} \,,
\end{eqnarray}
where
\begin{equation}\label{T-def-q}
\mathbf{R}_a{}^b := \mathbf{Z}_a\bar \mathbf{Z}^b- \cosh\left(\frac{q_a-q_b}{2}\right)\{ {\bm{\Phi}}, \bar{\bm{\Phi}} \}_a{}^b
\end{equation}
are the quantum counterparts of the quantities \p{T-def}.

The quantum supercharges \p{Q-q}, \p{bQ-q} form the superalgebra
\begin{eqnarray}\label{DB-QQ-q}
\{\mathbf{Q}, \mathbf{Q} \} &=&  \frac{1}{4}\sum\limits_{a\neq b}
\frac{ {\bm{\phi}}_a{}^b {\bm{\phi}}_b{}^a}{\sinh^2\Big({\displaystyle\frac{q_a-q_b}{2}}\Big)}\,\Big( \mathbf{F}_a-\mathbf{F}_b\Big) \,,
\\ [6pt]
\label{DB-bQbQ-q}
\{\bar \mathbf{Q} , \bar \mathbf{Q} \} &=&  \frac{1}{4}\sum\limits_{a\neq b}
\frac{ \bar{\bm{\phi}}_a{}^b \bar{\bm{\phi}}_b{}^a}{\sinh^2\Big({\displaystyle\frac{q_a-q_b}{2}}\Big)}\,\Big( \mathbf{F}_a-\mathbf{F}_b\Big) \,,
\\ [6pt]
\label{DB-QbQ-q}
\{ \mathbf{Q} , \bar \mathbf{Q} \} &=&  2\,\mathbf{H}  +\frac{1}{4}\sum\limits_{a\neq b}
\frac{ \bar{\bm{\phi}}_a{}^b {\bm{\phi}}_b{}^a}{\sinh^2\Big({\displaystyle\frac{q_a-q_b}{2}}\Big)}\,\Big( \mathbf{F}_a-\mathbf{F}_b\Big) \,,
\\ [6pt]
\label{DB-HQ-q}
[ \mathbf{Q}, \mathbf{H} ] &=&
-\frac{i}{8}\sum\limits_{a\neq b}
\frac{ \mathbf{R}_a{}^b {\bm{\phi}}_b{}^a}{\sinh^3\Big({\displaystyle\frac{q_a-q_b}{2}}\Big)}\,\Big( \mathbf{F}_a-\mathbf{F}_b\Big) \,,
\\ [6pt]
\label{DB-HbQ-q}
[ \bar\mathbf{Q}, \mathbf{H} ] &=&
-\frac{i}{8}\sum\limits_{a\neq b}
\frac{ \bar{\bm{\phi}}_a{}^b \mathbf{R}_b{}^a }{\sinh^3\Big({\displaystyle\frac{q_a-q_b}{2}}\Big)}\,\Big( \mathbf{F}_a-\mathbf{F}_b\Big) \,,
\end{eqnarray}
where the quantum Hamiltonian is defined by the expression
\begin{equation}\label{Ham-q}
\mathbf{H} = \frac12\,\sum_{a}\mathbf{p}_a \mathbf{p}_a +
\frac18\,\sum_{a\neq b} \frac{\mathbf{R}_a{}^b\mathbf{R}_b{}^a}{\sinh^2 \Big({\displaystyle\frac{q_a-q_b}{2}}\Big)} -
\frac18 \,{\rm Tr}\Big( \{{\bm{\Phi}}, \bar{\bm{\Phi}}\} \{{\bm{\Phi}},\bar{\bm{\Phi}}\}\Big)+
\frac{n(4n^2-1)}{24} \,.
\end{equation}
The operators
\begin{equation}\label{F-constr-q}
\mathbf{F}_a := \mathbf{R}_a{}^a -c= \mathbf{Z}_a\bar {\mathbf{Z}}^a- \{ {\bm{\Phi}}, \bar{\bm{\Phi}} \}_a{}^a - c \qquad\mbox{(no summation over $a$)}\,,
\end{equation}
that are on the right-hand sides of relations \p{DB-QQ-q}-\p{DB-HbQ-q}
are the quantum counterparts of the classical first class constraints \p{F-constr-d}.
We see that the quantum superalgebra \p{DB-QQ-q}-\p{DB-HbQ-q} is similar to the classical superalgebra \p{DB-QQ}-\p{DB-HbQ}.
Moreover, in the space of physical states $|\Psi\rangle$ that obey the conditions $\mathbf{F}_a |\Psi\rangle =0$,
the operators $\mathbf{Q}$, $\bar\mathbf{Q}$, $\mathbf{H}$ form the  $\mathcal{N}{=}\,2$ superalgebra.

In contrast to the classical supercharges \p{Q1}, \p{bQ1},
the quantum supercharges  \p{Q-q}, \p{bQ-q} have the following special property:
the first two terms
\begin{eqnarray}\label{Q-q-1}
\mathbb{Q} &:=&\sum\limits_{a} \mathbf{p}_a {\bm{\varphi}}_a  -\frac{i}{4}\sum\limits_{a\neq b}
\coth\Big({\displaystyle\frac{q_a-q_b}{2}}\Big)\Big({\bm{\varphi}}_a - {\bm{\varphi}}_b \Big)
\,,
\\
\label{bQ-q-1}
\bar{\mathbb{Q}} &:=&\sum\limits_{a} \mathbf{p}_a \bar{\bm{\varphi}}_a  + \frac{i}{4}\sum\limits_{a\neq b}
\coth\Big({\displaystyle\frac{q_a-q_b}{2}}\Big)\Big(\bar{\bm{\varphi}}_a - \bar{\bm{\varphi}}_b \Big)
\end{eqnarray}
in the supercharges  \p{Q-q}, \p{bQ-q} are separated and, moreover,
do not contain off-diagonal fermions ${\bm{\phi}}_a{}^b$, $\bar{\bm{\phi}}_a{}^b$.
The generators \p{Q-q-1} and \p{bQ-q-1} themselves form the $\mathcal{N}{=}\,2$ superalgebra
\begin{equation}\label{salg-1}
\{{\mathbb{Q}},\bar{\mathbb{Q}} \}= 2\,\mathbb{H}\,,\qquad \{\mathbb{Q},\mathbb{Q} \}=\{\bar{\mathbb{Q}},\bar{\mathbb{Q}} \}=0\,,
\end{equation}
where the Hamiltonian of such a ``truncated'' subsystem is given by the following expression:
\begin{equation}\label{H-q-1}
\mathbb{H}=\frac12 \sum\limits_{a} \mathbf{p}_a \mathbf{p}_a +
\frac{1}{8} \sum\limits_{a\neq b}
\frac{\Big(\bar{\bm{\varphi}}_a - \bar{\bm{\varphi}}_b \Big) \Big({\bm{\varphi}}_a - {\bm{\varphi}}_b \Big)}{\sinh^2\Big({\displaystyle\frac{q_a-q_b}{2}}\Big)} +
\frac{n(n^2-1)}{24}\,.
\end{equation}
Thus, the system with the generators \p{Q-q-1}, \p{bQ-q-1} and \p{H-q-1} is in fact the $\mathcal{N}{=}2$ special extension
of the hyperbolic Calogero-Sutherland system with a fixed value of the coupling constant.
A similar $\mathcal{N}{=}2$ Calogero-Sutherland system in the trigonometric case 
was studied in \cite{DeLaMa}. Note, the system with Hamiltonian \p{H-q-1} is similar to the system \cite{DeLaMa} 
with the coupling constant equal to unity. 

Thus, we have obtained an important conclusion regarding the supercharges of the considered Calogero-Sutherland system.
In the classical charges \p{Q}, all non-trivial terms are proportional to the off-diagonal fermions $\phi_a{}^b$, $\bar\phi_a{}^b$
and pass away when these fermions disappear.
That is, in the classical case, the presence of off-diagonal fermions is the key point in constructing the $\mathcal{N}{=}\,2$ extension
of the Calogero-Sutherland system as a gauged superfield matrix model.
In the quantum case, the physical space of the model contains the subsector without off-diagonal fermions,
which is invariant with respect to the $\mathcal{N}{=}\,2$ supersymmetry. The supersymmetry generators in this subsector are
presented in  \p{Q-q-1}, \p{bQ-q-1}, \p{H-q-1}.

\setcounter{equation}{0}
\section{Concluding remarks and outlook}

In this paper, a classical and quantum description
of the $\mathcal{N}{=}\,2$ supersymmetric multi-particle hyperbolic Calogero-Sutherland system is presented,
which was obtained from the matrix superfield model by the gauging procedure.
Explicit expressions are obtained for the classical and quantum generators of the $\mathcal{N}{=}\,2$ supersymmetry,
corresponding to the hyperbolic Calogero-Sutherland system considered here.

In the fully matrix system, the supercharges \p{Q-matrix} have a simple form, but a system like this has a large number of gauge degrees of freedom corresponding to the $u(n)$ first class constraints \p{F-constr}, \p{F-constr1}.
In the reduced system without off-diagonal even matrix variables, the supercharges \p{Q} (or \p{Q1}, \p{bQ1})
have the Calogero-like form.
The $\mathcal{N}{=}\,2$ supersymmetry algebra is closed in this case only on the shell of the residual $[u(1)]^n$ first class constraints \p{F-constr-d}.
It is emphasized that non-trivial terms in the classical supercharges \p{Q} vanish when off-diagonal odd variables disappear.
This is not so in the quantum case, where the obtained quantum supersymmetry generators \p{Q-q}, \p{bQ-q}
contain the terms \p{Q-q-1}, \p{bQ-q-1} that are independent of off-diagonal fermions and form
the $\mathcal{N}{=}\,2$ supersymmetry algebra.
In addition, the Lax pair \p{L-matr}, \p{M-matr} is found for the system under consideration
and the set of the conserved quantities \p{Ik-def}, \p{Ik-f-def}, \p{J-f-def}, \p{Ik-f2-def} is defined.
An analysis of the integrability of the system considered here will be the subject of the next paper.

In the next publications, it is planned to study
the $\mathcal{N}{=}\,4$ supersymmetric hyperbolic Calogero-Sutherland system constructed in \cite{FIL19}.
In contrast to the $\mathcal{N}{=}\,2$ case discussed in this paper, the $\mathcal{N}{=}\,4$ supersymmetric generalization
has the $U(2)$ spin hyperbolic Calogero-Sutherland system as a bosonic core.
An interesting question here is to understand the role of spin variables in systems of this type
and to make possible the separation of the invariant subsector without off-diagonal odd variables in the quantum case.

One more interesting problem is to construct the $\mathcal{N}{=}\,2$ and $\mathcal{N}{=}\,4$ supersymmetric generalizations
of the trigonometric Calogero-Sutherland system by the gauging procedure of some matrix systems.

\smallskip
\section*{Acknowledgements}
I  would  like  to  thank  Alexei Isaev, Evgeny Ivanov  and  Sergey Krivonos  for useful discussions.
This work was supported by the Russian Science Foundation, grant no.\,16-12-10306.


\end{document}